\documentclass[letter]{aa} % for the letters 
\usepackage{natbib,twoopt,xcolor}
\usepackage[breaklinks=true]{hyperref} %% to avoid \citeads line fills
\bibpunct{(}{)}{;}{a}{}{,}             %% natbib format for A&A and ApJ
\makeatletter
  \newcommandtwoopt{\citeads}[3][][]{\href{http://adsabs.harvard.edu/abs/#3}%
    {\def\hyper@linkstart##1##2{}%
     \let\hyper@linkend\@empty\citealp[#1][#2]{#3}}}
  \newcommandtwoopt{\citepads}[3][][]{\href{http://adsabs.harvard.edu/abs/#3}%
    {\def\hyper@linkstart##1##2{}%
     \let\hyper@linkend\@empty\citep[#1][#2]{#3}}}
  \newcommandtwoopt{\citetads}[3][][]{\href{http://adsabs.harvard.edu/abs/#3}%
    {\def\hyper@linkstart##1##2{}%
     \let\hyper@linkend\@empty\citet[#1][#2]{#3}}}
  \newcommandtwoopt{\citeyearads}[3][][]%
    {\href{http://adsabs.harvard.edu/abs/#3}
    {\def\hyper@linkstart##1##2{}%
     \let\hyper@linkend\@empty\citeyear[#1][#2]{#3}}}
\makeatother

\usepackage{graphicx}
\usepackage{txfonts}
\makeatletter
\renewcommand*\aa@pageof{, page \thepage{} of \pageref*{LastPage}}
\makeatother

\begin{document}

   \title{Similar additional frequency patterns on fundamental and overtone mode RR Lyrae stars showing $f_{68}$ frequencies}
%   \subtitle{}
\titlerunning{Additional frequency patterns on fundamental and overtone mode RR Lyrae stars}
\authorrunning{Benk\H{o} \& Kov\'acs}

   \author{J\'ozsef M. Benk\H{o}
          \inst{1}
          \and
          G\'abor B. Kov\'acs\inst{2,1}
          }

   \institute{Konkoly Observatory, HUN-REN Research Centre for Astronomy and Earth Sciences; 
              MTA Centre of Excellence,        \\ 
              Konkoly Thege u. 15-17., 1121 Budapest, Hungary; 
              \email{benko@konkoly.hu}
%         \and
%             MTA CSFK Lendület Near-Field Cosmology Group
         \and
             Department of Astronomy, Institute of Geography and Earth Sciences, ELTE E\"otv\"os Lor\'and University, P\'azm\'any P\'eter s\'et\'any 1/A, H-1117 Budapest, Hungary \\
             }

   \date{Received Sep 25, 2023; accepted Nov 20, 2023}

% \abstract{}{}{}{}{} 
% 5 {} token are mandatory
 
  \abstract
   {Up to now, it seemed that the additional frequencies in the fundamental mode (RRab) and in the overtone mode pulsating (RRc and RRd) RR Lyrae stars have different nature.
   RRab stars show frequencies associated with periodic doubling, as well as frequencies at the first and second radial overtones, and linear combinations of these. RRc stars show frequencies with specific ratios ($f_1/f_x\sim$0.61, or $\sim$0.63) which are explained by non-radial modes and frequencies with $f_x/f_1\sim0.68$ ratio which have no currently accepted explanation.}
   {
   To search for similarities in spectral content, we compared the Fourier spectra of the recently published TESS and K2 data of RRc stars with the spectra of Kepler RRab stars that do not show the Blazhko effect but contain additional frequencies.
   }
   {
   The time series data have been analysed using standard Fourier methods, and the possibility of the excitation of the second radial overtone mode in RRab stars has also been tested using numerical hydrodynamical codes.
   }
   {We show that the additional frequencies  appear  in non-Blazhko RRab stars at the position of the second radial overtone mode, and the pattern they create, is very similar to that caused by the additional frequencies with the period ratio $\sim0.68$ in RRc stars. 
   The former explanation of the additional frequencies of these RRab stars by a second radial overtone is unlikely.
   }
   {}

   \keywords{Stars: oscillations --
                Stars: variables: RR Lyrae --
                Methods: data analysis --
                Techniques: photometric --
                Space vehicles
               }

   \maketitle
%
%-------------------------------------------------------------------

\section{Introduction}

   In the last two decades, the view we look at stars pulsating in the classical 
   instability strip (e.g. Cepheides, RR Lyrae stars) has changed fundamentally. 
   Thanks to photometric space missions MOST \citep{Walker2003}, 
   CoRoT \citep{Baglin2006}, Kepler/K2 \citep{Borucki2010, K22014}, and TESS \citep{Ricker2015};
   as well as the ground-based survey OGLE \citep{Udalski2015},
   a number of millimagnitude level brightness variations have been 
   detected in these stars. The frequencies that describe these brightness changes are usually
   called additional frequencies.
   Since this paper is about RR Lyrae stars, other groups of the instability strip showing similar additional frequencies (e.g. classical Cepheids, anomalous Cepheids, Pop II Cepheids, etc) will not be discussed here. 
   We do not discuss here also the frequency patterns caused by the Blazhko effect (long-period amplitude and phase variation of the light curve -- see for a review e.g. \citealt{Kovacs_Blazhko2009, Smolec_Blazhko}).
   Some papers have already reviewed the additional
   frequencies of RR Lyrae stars (see e.g. \citealt{Molnar2017, Plachy_Szabo2021}),  we will only briefly summarise the story.

   The first frequency that did not fit into the previously known frequency structure of RR Lyrae stars was detected by \citet{Gruberbauer2007} from the MOST data of the double mode (RRd) star AQ Leo. Later, similar frequencies have been found in other RRd stars, and many RRc stars which pulsate purely in the overtone mode \citep{Olech2009, Chadid2012, Szabo2014, Moskalik2015}. 
   After analysing the large amount of RR Lyrae stars of the OGLE survey it became clear that these frequencies are located in three horizontal roughly parallel sequences on the Petersen diagram, where the frequency ratio $f_1/f_x$ of each sequence is 0.613, 0.622, and 0.631, respectively \citep{Netzel2015, Netzel2019}. (From now on, these frequencies will be referred to as  $f_{61}$, $f_{62}$ and $f_{63}$.)
   In the meantime, a theoretical explanation has been suggested by \citet{Dziembowski2016}: the frequencies $f_{61}$ and $f_{63}$ belong to non-radial modes $\ell=9$ and $\ell=8$, respectively, while $f_{62}$ is a linear combination frequency.
   The frequencies of the modes themselves are $0.5f_{61}$ and $0.5f_{63}$.
   This picture was confirmed by further theoretical and observational studies \citep{Netzel2021, Netzel2022, Molnar2022, Benko2023, Netzel2023}. 

\begin{table*}
\caption{\label{Tab:param_RRab}Non-Blazhko Kepler/K2 RRab stars showing $f_2$ additional frequencies
}
\centering
\begin{tabular}{llccccrll}
\hline\hline
\noalign{\smallskip}
Star& $P_0$ & $f_0/f_2$ & $f^{0}_{68}/f_0$ & $A^{0}_{68}/A_{2}$ & $K_{\mathrm p}$ & $\Delta t$ & [Fe/H]  & Sources\\
    & (d) & &  & & (mag) &  (d)   &  & \\
\hline
\noalign{\smallskip}
KIC 9658012    &   0.533195  & 0.5926 &  0.6874 & 0.92 & 16.001 &  684.14   & $-1.28\pm0.14$  & 1,3      \\
V894 Cyg       &   0.5713865  & 0.5934 &  0.6850 & 1.13 & 13.293 & 1470.46  & $-1.66\pm0.12$  & 1,3      \\
V346 Lyr       &   0.5768270 & 0.5931 &  0.6860 &  1.50 & 16.421 &  1459.49 & $-1.82\pm0.03$\tablefootmark{$\ast$} & 1,3    \\
V1510 Cyg      &   0.5811426 & 0.5936 & 0.6848 &  1.54 & 14.494 &  1459.49 & $-1.80\pm0.03$\tablefootmark{$\ast$}  & 1,3   \\
%V350 Lyr       &   0.5942378 & 0.5925 & 15.696   & $-1.83\pm0.12$      &     \\
%KIC 7021124    &   0.6224813 & 0.5928 & 0.6869 & 1.16 & 13.550 &  1459.47 & \\
EPIC 60018644  &   0.64504  & 0.5853 & 0.6876 & 1.79 &12.199 &  8.97 & $-1.88\pm0.1$\tablefootmark{$\ast$}   &  2 \\
\hline
\end{tabular}
\tablefoot{
(1) \citet{Benko2019};
(2) \citet{Molnar2015};
(3) \citet{Nemec2013} \\
\tablefoottext{$\ast$}{Metallicity calculated from photometry.}
}
\end{table*}  

   A further group of additional frequencies have been discovered in RRc stars \citep{Netzel2015, Moskalik2015}. The ratio of these frequencies are $f_1/f_x \approx 1.46$ or  $f_x/f_1 \approx 0.68$. (Hereafter as $f_{68}$ frequencies.) The $f_{68}$ frequencies are not only lower than the dominant $f_1$, but also lower than the fundamental frequencies of the stars in question. They can appear alone or in combination with the aforementioned $f_{61}$, $f_{63}$ frequencies.  Currently, there is no theoretical explanation for their presence. 

   RRab stars may also contain additional frequencies, but different from the ones we have discussed. 
   The first additional frequencies on an RRab star were found in the CoRoT data of V1127 Aql,
   which also shows a strong Blazhko effect \citep{Chadid2010}.
   The Kepler space telescope data of other Blazhko RRab stars has made it clear that these 
   frequencies are associated with the period doubling (PD) phenomenon \citep{Kolenberg2010, Szabo2010}. 
   In the case of RR Lyrae light curves, PD appears as an alternation of cycles with smaller and larger amplitudes causing half-integer frequencies ($f_0/2$, $3f_0/2$, etc.) in the Fourier spectra, where $f_0$ is the frequency of the radial fundamental mode. Hydrodynamic calculations have shown that for RR Lyrae stars the PD is caused by a 9:2 resonance of the fundamental mode and the ninth radial overtone \citep{Kollath2011}. So far, PD has not been detected in either RRc or RRd stars, and among RRab stars, it has only been found in stars that also show the Blazhko effect.
   The theoretical works have also suggested that the PD and the Blazhko effect may be related to each other \citep{Buchler2011, Kollath2011}. 

   In addition to the PD frequencies, some RRab stars show other low amplitude frequencies near the first and/or second radial overtones \citep{Benko2010}. (These frequencies will be denoted by $f_1$ and $f_2$, respectively.) They were later identified also in ground-based observations \citep{Sodor2012, Smolec_Prudil2016, Zalian2016}, and in the CoRoT, Kepler/K2 and TESS samples \citep{Szabo2014, Benko2014, Benko2016, Molnar2015, Molnar2022}. 
   Hydrodynamic calculations have been used to produce models in which the fundamental mode, 
   the first and the ninth overtone appear simultaneously in a triple resonance state. This provides a possible theoretical explanation for stars exhibiting both PD and $f_1$ \citep{Molnar_Kollath2012}. However, despite an intensive search, similar models could not be found 
   for the $f_2$ cases. Using linear models, \citet{Soszynski2016aRRd} showed that, for sufficiently high metallicity, parametric resonance can generate states in which the fundamental, first and second overtones are excited simultaneously. These models could be suitable for a fraction of observed stars only. However, \citet{Dziembowski1999} have shown that the non-radial modes associated with $\ell=1$ are most likely to be excited near the radial modes ($\ell=0$), i.e. both $f_1$ and $f_2$ additional frequencies can be those of such non-radial modes.

   The additional frequency content of the fundamental and overtone mode RR Lyrae stars seemed to be disjoint:
   in RRab stars, patterns belonging to PD, $f_1$ and $f_2$ frequencies, while RRc and RRd stars, patterns belonging to the frequencies $f_{61}$, $f_{63}$ and $f_{68}$ may appear. 
   
\section{The sample}
     
    The Blazhko effect produces characteristic patterns of many peaks in the Fourier spectra. The additional frequencies also have distinctive patterns with many harmonics and linear combinations with the main pulsation frequency. To ensure that the two structures do not interfere with each other, we focus on the additional frequencies of the non-Blazhko stars. 

    \subsection{RRab stars with \texorpdfstring{$f_2$}\ \ frequencies}

    Few RRab stars have been known that are unlikely to show the Blazhko effect, but additional frequencies do appear in their Fourier spectra. Long time series are required to rule out the Blazhko effect, and precise measurements are needed to detect additional frequencies in millimagnitude scales. Not even all space photometric data fulfil both conditions.

    The first two non-Blazhko RRab stars to detect additional frequencies were V350 Lyr and  KIC 7021124 \citep{Benko2010, Nemec2011}. The frequency of the second radial overtone mode $f_2$ and its linear combination were identified for both stars. Later, however, a very weak Blazhko effect was detected for both stars \citep{Benko2015} so these stars were not analysed further.
    \citet{Molnar2015} found that \object{EPIC 60018644} may also be a non-Blazhko star with an additional frequency pattern. Since this was obtained from the commissioning phase (C0) data of K2, with a length of 8.9 days only, the Blazhko effect cannot be ruled out. The highest amplitude additional frequency is shorter than the fundamental mode frequency and is therefore referred to as a g-mode frequency $f_g$. Harmonics and linear combinations of this frequency also appear in the Fourier spectrum.  

    A systematic study of the Kepler non-Blazhko RRab stars has found further stars with additional frequency structures: \object{V1510 Cyg}, \object{V346 Lyr}, \object{V894 Cyg} and \object{KIC 9658012}.  In all of them, the $f_2$ frequency and its combinations were identified \citep{Benko2019}. 
    Due to the short data series, the signal-to-noise ratio of the frequencies of EPIC 60018644 is much lower than in the spectra of the Kepler stars, but their similar patterns are still evident (see figure 2 in \citealt{Benko2015} and figure 11 in \citealt{Molnar2015}, respectively).
    Some important parameters of our RRab sample are summarised in Tab.~\ref{Tab:param_RRab}. The precision of the numeric values (where not specifically shown), is indicated by the number of significant digits plus one.  This convention has also been applied to subsequent tables.

    \subsection{RRc stars showing \texorpdfstring{$f_{68}$}\ \ frequencies}

    As in the case of RRab stars, we intend to separate the different frequency patterns as much as possible. To do this, only those non-Blazhko RRc stars are considered, which  do not show the $f_{61}/f_{63}$ frequencies as well. The CoRoT and the original Kepler samples do not contain such `clear' cases, however, the works on TESS and K2 RRc stars identified significant number of such stars \citep{Benko2023, Netzel2023}.
    From the TESS sample 23 stars were selected that contain both the $f_{68}$ frequency and at least one linear combination of $f_{68}$ and $f_1$. Uncertain and blend cases, as well as stars where only the $f_{68}$ frequency is significant, were excluded. Some parameters of the sample are shown in Tab.~\ref{Tab:param_RRc_TESS}. 
    We identified nine of the K2 RRc stars in the work of \citet{Netzel2023} that met our criteria (see them in Tab.~\ref{Tab:param_RRc_K2}).

    The number of RRc stars is higher than that of the RRab sample.    
    It should be noted, however, that only two stars (Gaia DR2 5482545510194122112 and Gaia DR2 4628067852624828672) in the TESS continuous viewing zone have a data set of comparable length to the four years of Kepler data.

\section{The result}

   \begin{figure}[h!]
   \centering
   \includegraphics[width=8.2cm]{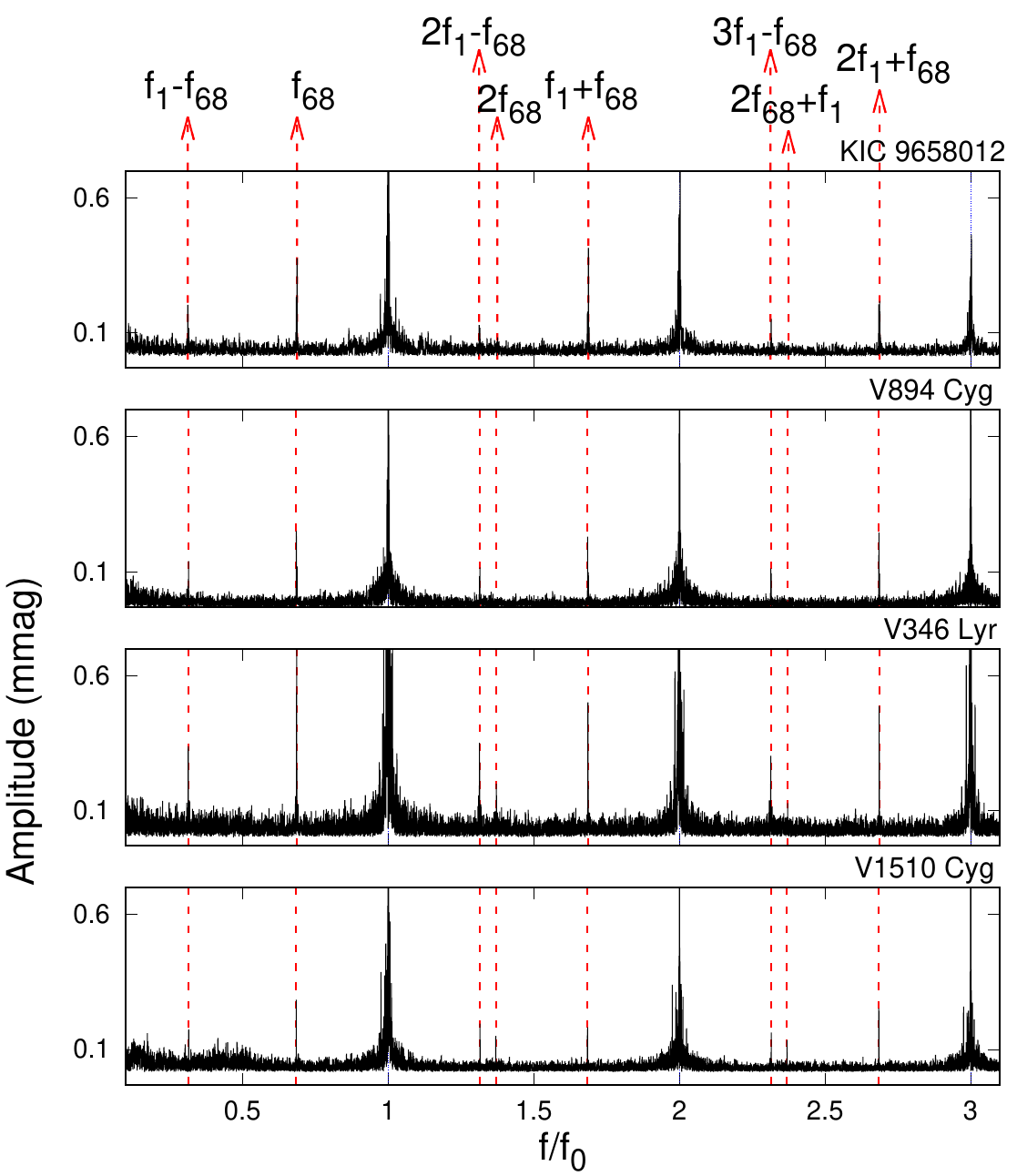}
   \includegraphics[width=8.2cm]{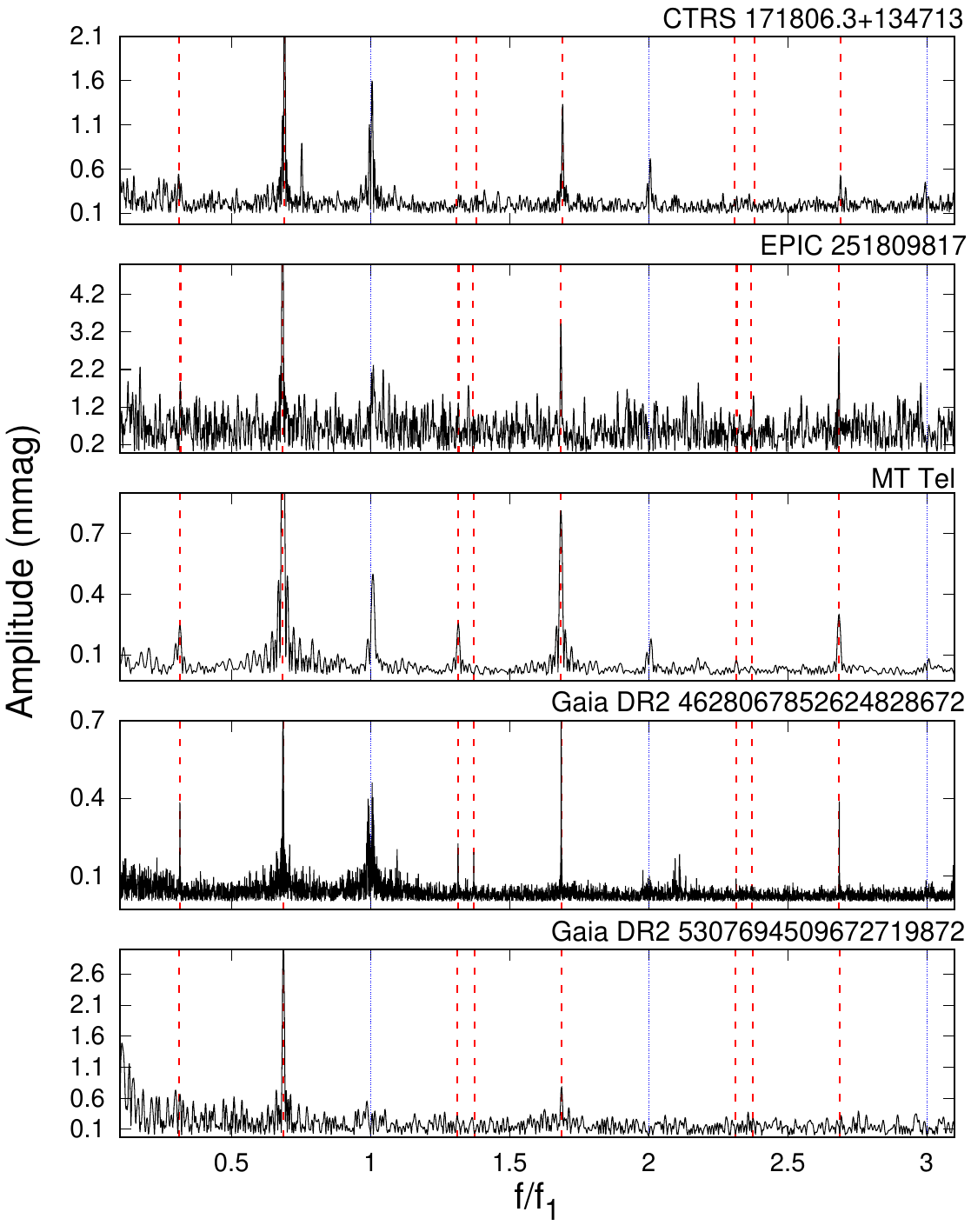}
      \caption{Residual spectra of non-Blazhko RRab and RRc stars after the 
      dominant pulsation frequency and its harmonics are pre-whitened from the data.
      Top  four panels show the Kepler RRab stars of Table~\ref{Tab:param_RRab} while bottom five panels
      illustrates the spectra of RRc stars. The blue dotted and red dashed lines show the position of the (pre-whitened) harmonics and the possible position of the additional frequencies, respectively.
      The identification on top belongs to the RRc stars containing $f_{68}$ frequencies.
              }
         \label{fig:spectra}
   \end{figure}

Let us compare the frequency content of the RRab and RRc samples described above.  
The Fourier spectra of Kepler RRab stars have already been prepared in our previous works \citep{Benko2015, Benko2019}
and we have adopted them here.
The underlying photometric time series are available for download\footnote{{\url{https://konkoly.hu/KIK/}}}. 
The residual spectra of the Kepler stars in Table~\ref{Tab:param_RRab} are shown in the top four panels 
of Fig.~\ref{fig:spectra}. These spectra were produced from the data series from which the main period and all its significant harmonics were pre-whitened. The spectrum of the K2 star EPIC 60018644 is given in figure 11 of the paper in \citet{Molnar2015} that is not repeated here. As shown in the Table~\ref{Tab:param_RRab}, the period increases from top to bottom. For better comparison, the relative frequencies $f/f_0$ are plotted on the horizontal scales.  
 
The Fourier spectra of TESS  RRc stars are taken from the work of \citet{Benko2023}. For the K2 RRc stars in Table~\ref{Tab:param_RRc_K2}, following \citet{Netzel2023}, we downloaded the Automated Extended Aperture Photometry data \citep{Bodi2022} than the spectra were calculated by using the Period04 program \citep{Lenz2005}.
The lower five panels of Fig.~\ref{fig:spectra} show residual spectra of some RRc stars.
Stars with different parameters have been selected to represent different observed cases. 
We include the spectra of both the brightest (MT Tel, $T_\mathrm{mag}=8.757$~mag) and the faintest star (EPIC 251809817, $K_{\mathrm p}=19.008$~mag) from our sample\footnote{Because of the different sensitivity of the two detectors, the $K_{\mathrm {p}}$ and TESS magnitudes are slightly different, but this is not relevant for our purposes.}. Similarly, the spectra of the shortest
and longest period stars are shown: CTRS 171806.3+134713, $P_1=0.27933$~d, and Gaia DR2 5307694509672719872,  $P_1=0.376984$~d. Gaia DR2 4628067852624828672 is included here as the RRc star with the two longest TESS data sets, with a spectrum that not only has an outstandingly good signal-to-noise ratio, but also the richest additional frequency pattern. Relative frequencies ($f/f_1$) are shown on the horizontal scale.

As can be seen in Fig.~\ref{fig:spectra}, the additional frequencies are located at the same positions on this relative frequency scale for both RRab and RRc stars. At the top of the figure, we have assigned to these positions the identifications adopted for RRc stars. For RRab stars, the frequency at the position of $f_0+f_{68}$ is usually considered as primary and identified as the second overtone frequency $f_2$. 
More quantitatively, the average period ratio of the $f_{68}/f_1$ for TESS and K2 RRc samples are $0.684\pm0.004$ and $0.685\pm0.003$, respectively, while for RRab stars the ratio of $(f_2-f_0)/f_0$ is $0.686\pm0.001$.
That is, all three samples really show the same ratios within 1$\sigma$ error. 
This raises the possibility that the frequencies with a ratio of $\sim0.68$ are the primary 
ones for RRab stars too. From now on these frequencies are denoted as $f^0_{68}=(f_2-f_0)$.

\section{Discussion}

\begin{table}[]
    \centering
    \caption{Modelling input parameters of the studied stars.}
    \begin{tabular}{lllll}
    \hline
    \hline
    \noalign{\smallskip}
    Star & $M$\tablefootmark{(a)}        & $T_{\rm eff}$ & $L_{\rm bol}$ & $Z$\tablefootmark{(b)} \\
         &($M_\odot$) & (K) & ($L_\odot$)&  \\
    \hline
    \noalign{\smallskip}
    KIC 9658012     & $0.5281$ & $6464\pm100$\tablefootmark{$\ast$} & $36\pm5$ & $0.0011$\\
    V894 Cyg       & $0.63089$ & $6330\pm100$\tablefootmark{$\ast$} & $42\pm2$ & $0.0004$\\
    V346 Lyr       & $0.72084$ & $6205\pm100$\tablefootmark{$\ast$} & $43\pm7$ & $0.0003$\\
    V1510 Cyg     &  $0.66043$ & $6393\pm200$\tablefootmark{$\ast\ast$}  & $46\pm4$ & $0.0003$\\
%    KIC 7021124    &  $ 0.6883$ & $6535\pm200$\tablefootmark{$\ast\ast$} & $57\pm7$ & $0.0001$\\
    EPIC 60018644  &  $0.72600$ & $6333\pm200$\tablefootmark{$\ast\ast$} & $54\pm8$ & $0.0003$\\
    \hline
    \end{tabular}
    \tablefoot{
    \tablefoottext{a}{Interpolated on linear non-adiabatic period-mass grids.}\\
    \tablefoottext{b}{Hydrogen content is assumed to be $X=0.754$.}\\
    \tablefoottext{$\ast$}{Temperature calculation are based on Gaia mean-colors, [Fe/H] and spectroscopy.}\\
    \tablefoottext{$\ast\ast$}{Temperature calculation is only based on Gaia mean-colors and [Fe/H].}
    }
    \label{tab:model_input}
\end{table}

Let us then investigate the Fourier spectra in the framework of the above paradigm, 
i.e. assuming that for both RRab and RRc stars the $f_{68}$ and $f^0_{68}$ are the primary frequencies.

In all cases the stars show the linear combination of $f_{68}$ and $f^0_{68}$ as 
frequencies $f_{59}=f_{68}+f_1$ and $f^0_{59}=f^0_{68}+f_0$. The $f_1-f_{68}$ (and $f_0-f^0_{68}$) combinations also appears on many stars but not always.  It seems to be generally true that the amplitude of 
combinations of differences is smaller than that of combinations of sums, that is e.g. $A(kf_1+jf_{68}) > A(kf_1-jf_{68})$, where $k$, $j$ are positive integers and $k\ge j$. Fewer linear combinations with higher harmonics appear in the spectra of RRc stars. This is possibly because these stars have much more sinusoidal light curves than RRab stars, i.e. the higher harmonics carry much less power, and therefore, the linear combinations with these harmonics drop below the significance level very quickly.

It is known that $f_{68}$ frequencies are rather coherent \citep{Netzel2019, Benko2023}.
Using amplitude and phase variation calculation tool of Period04 each  RRab time series was cut into 10 and 7-day slices, and then determined the amplitudes and phases of $f^0_{68}$
for each slice. For all RRab stars, both amplitude and phase found to be constant within the estimated error. This behaviour is an another similarity between the $f_{68}$ and $f^0_{68}$ frequencies.

The amplitude of the $f_{68}$ frequencies, and with the exception of KIC 9658012 the $f^0_{68}$ frequencies, 
are larger than those of the $f_{59}$ and $f^0_{59}$ frequencies (see column 5 in Table~\ref{Tab:param_RRab} and cols 4 in Tables~\ref{Tab:param_RRc_TESS} and \ref{Tab:param_RRc_K2}). 
It is known that linear combinations of 
non-radial pulsation frequencies can have larger amplitudes than the amplitudes of the parent frequencies that form the combination  (e.g. \citealt{Balona2013}).
The theoretical explanation of the phenomenon was given by \citet{Kurtz2015}. 
The key is that the observed amplitude of a linear combination depends on the spherical harmonics of the component modes (see eq. (7) and formulae (8) and (9) in \citealt{Kurtz2015}).
However, in the case of two radial pulsation components the spherical harmonics became constants, i.e. the amplitudes of the combination peaks must always be smaller than those of the components. 
Applying this to our RRab stars, in four of the five stars in our sample, $f^0_{59}$ cannot be a radial mode frequency, since it has a smaller amplitude than the amplitude of its linear combination with $f_0$ (see col. 5 in Table~\ref{Tab:param_RRab})\footnote{
The amplitude ratios given in Table~\ref{Tab:param_RRab} are accurate to at least a few percents, i.e. it is not possible to reverse the ratios due to observation noise.}, and thus $f^0_{59}$ cannot be identified with $f_2$. 

By comparing the amplitude ratios of the $f_{68}$ and $f_{59}$ (or $f^0_{68}$ and $f^0_{59}$) frequencies: $A_{68}/A_{59}$ (or $A^0_{68}/A_2$) 
in Tables~\ref{Tab:param_RRab}-\ref{Tab:param_RRc_K2}, we see that this ratio for
RRc stars tends to be higher ($\sim$2.9) than for RRab stars ($\sim$1.4). 
Harmonics of the $f^0_{68}$ or $f_{68}$ frequencies also appear in some cases (see e.g. V346 Lyr, V1510 Cyg 
or Gaia DR2 4628067852624828672).

We made linear non-adiabatic (LNA) calculations to identify the parameter range in which second overtone pulsation is excited, and also how close are the RRab stars of our sample to this region. This calculation was performed with the Budapest-Florida code \cite{Yecko1998}, using the RR Lyrae convective parameter set of \citet{KovacsGB2023b}. The input parameters for the calculation are:
the pulsating mass ($M$), the effective temperature ($T_{\rm eff}$), the bolometric luminosity ($L_{\rm bol}$), the hydrogen ($X$) and metal ($Z$) mass fractions.
$M\in[0.3,0.9]$ with  0.05 $M_\odot$ steps, $L\in[35,80]$ with  2 $L_\odot$ steps and $T_{\rm eff}\in[6000,7400]$~K with $50$~K steps and for three metallicites: $Z=0.0001$, 0.0003, and 0.0011. While $X$ was chosen by assuming primordial Helium content \citep[$Y=0.2465$][]{Helium}.

To place the studied stars in this parameter grid,
 the effective temperatures were determined by comparing the spectroscopic $T_{\rm eff}$  of \cite{Nemec2013} to the Gaia and 2MASS colour-$T_{\rm eff}$ relations\footnote{We used the \texttt{Colte} module for this purpose} of \cite{Casagrande2021} and the statistical $T_{\rm eff}-[{\rm Fe}/{\rm H} ]$ relation of \cite{Jurcsik1998}. To get the bolometric luminosities, we calculated absolute $K$ and $V$ magnitudes by period-luminosity and period-Wasenheit ($K$, $V-K$ and $J-K$) relations \citep{Cusano2021} using extinction data of \cite{Schlafly2011}, and bolometric corrections of \cite{Bessel1998}. Pulsation mass ($M$) were determined through interpolating with the observed period on the linear non-adiabatic (LNA) model grid using the other parameters. This method gives similar results as the PLMTZ relations of \cite{Marconi2015}, while ensures that a non-linear calculation would have periods to the linear ones as close as possible. These parameters can be found in Table~\ref{tab:model_input}.

\begin{figure}
    \centering
    \includegraphics[width=\columnwidth]{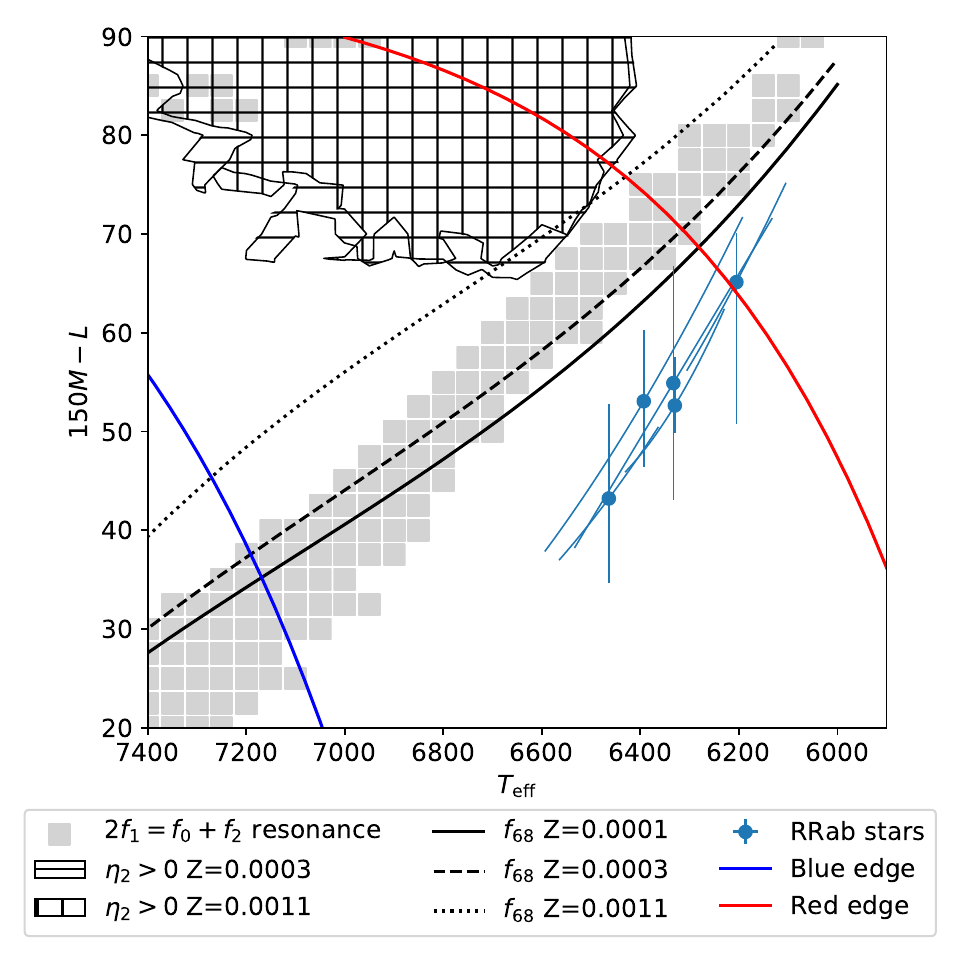}
    \caption{Position of the observed stars on the $T$--$(150M-L)$ plane. Light blue circles with error-bars show the stars from Table~\ref{tab:model_input}. The hatched area shows where the second overtone is excited. The different patterns indicate the different metallicites.  The dotted, dashed and solid black curves correspond to those models that have the ratio $(f_2-f_0)/f_0=0.686$ for the linear grids with metallicities $Z=0.0001$, $Z=0.003$ and $Z=0.0011$ respectively. The blue and the red solid curves are the  first overtone blue and the fundamental red edges of the instability strip with $M=0.65 M_\odot$ for reference as given by \cite{Marconi2015}. Grey squares denotes models with $|2f_1 - f_0 - f_2 | < 0.005$ resonance. } 
    \label{fig:ml-t}
\end{figure}

We looked for cases where the linear growth rate of the second overtone $\eta_2>0$ i.e. it is an excited mode. We show these results on the $150M-L$ \citep{Kollath2011} vs. $T_{\rm eff}$ plane in Fig.~\ref{fig:ml-t}. In this $150M-L$ projection the more luminous stars are in the negative direction, while the more massive stars are in the positive direction. We can see in Fig.~\ref{fig:ml-t} that to have excitation for the second overtone the star needs relatively large mass and/or low luminosity ($150M-L > 66$), while the temperature region is also higher then those of the studied stars. {We note that this region is different than the "typical" 2O pulsation region of RR Lyraes, that is shifted towards higher temperatures.} The frequency ratio of the studied stars in the models are $(f_2-f_0)/f_0 \approx 0.73 \pm 0.03$. Which is not consistent with the observed $f_{68}$. 

Another possibility to produce additional frequencies are non-linear effects e.g. resonant mode pulsations \citep{Kollath2011}. Although a full mode selection analysis needs large amount of non-linear calculations which is unfeasible, we can check one possible resonant region for $f_{68}$ signals: \cite{Soszynski2016aRRd} suggested the $2f_{1O} = f_{F} + f_{2O}$ resonance as  explanation of anomalous RRd stars. We show the models of these resonances in Fig.~\ref{fig:ml-t} with grey squares, where $|2f_{1O} - f_{F}- f_{2O}| < 0.005$. One can see, that indeed this resonance region crosses $f_{68}$ regions for lower metallicities, although the sample stars are not in this region.

\section{Conclusions}

We have shown that the additional frequencies on non-Blazhko RRab stars produce Fourier 
patterns quite similar to the $f_{68}$ frequencies known for RRc stars. 
The former explanation for these structures of RRab stars with the frequencies 
of the second radial overtone modes $f_2$ is unlikely since (i)
the amplitudes of the linear combination frequencies of radial modes cannot be higher than the amplitudes of the component frequencies, and (ii) the model calculations for all stars show that their measured physical parameters are very far from the range that allows a second overtone pulsation. The additional pattern of  some RRab and RRc stars has been successfully linked, unfortunately with a frequency that is not yet explained, nor even known to be pulsation in origin.

\begin{acknowledgements}
      This paper includes data collected by the Kepler and TESS missions. Funding for the missions is 
      provided by the NASA Science Mission Directorate. The research was partially supported by the ‘SeismoLab’ KKP-137523 \'Elvonal and NN-129075 grants of the Hungarian Research, Development and Innovation Office 
      (NKFIH).
\end{acknowledgements}


\bibliographystyle{aa}
\bibliography{Benko+Kovacs_aa_letter}

\listofobjects

\begin{appendix}

\section{}
\begin{table*}
\caption{\label{Tab:param_RRc_TESS}Non-Blazhko TESS RRc stars showing $f_{68}$ additional frequencies.
}
\centering
\begin{tabular}{lllllr}
\hline\hline
\noalign{\smallskip}
 Star & $P_1$ & $f_{68}/f_1$ & $A_{68}/A_{59}$ & $T_{\mathrm{mag}}$ & $\Delta t$  \\
 & (d) &  & & (mag) &    (d)  \\
\hline
\noalign{\smallskip}
	CRTS J171806.3+134713	&	0.279329	& 0.6905 & 2.30	& 13.333 &	51.46 \\ 
	MV Tel	&	0.303292& 0.6813	 	&	2.71 & 13.668	&	27.85 \\
	IS Com	&	0.314644	& 0.6855	& 2.38	& 13.570	&	26.75 \\
	ROTSE1 J161226.41+323225.5	&	0.315523	& 0.6866	& 1.86	& 13.681	&	53.48 \\
	Gaia DR2 4628067852624828672	&	0.316098	& 0.6854	& 2.79	& 13.081	&	243.12 \\
	MT Tel	&	0.316897	& 0.6851	& 2.09	& $\phantom{0}$8.757	&	28.35 \\
	Gaia DR2 4703977585650108032	&	0.317785	& 0.6851	&	2.43 & 13.872	&	56.12 \\
	BG Boo	&	0.320073	& 0.6853	&	3.44 & 13.271	&	26.69 \\ %
	GSC 02626-00896	&	0.322674	& 0.6852	& 2.64	& 12.654	&	24.83 \\
	AO Tuc	&	0.333227	& 0.6852	& 3.92	& 10.949	&	56.17 \\ %
  Gaia DR2 5482545510194122112	&	0.337873	& 0.6859	&	5.24 & 12.282	&	243.13 \\
	Gaia DR2 1844337262749036928	&	0.338736	& 0.6849	&	3.16& 13.011	&	25.96 \\
	Gaia DR2 6558066176505330304	&	0.341317	& 0.6843	&	2.40 & 12.604	&	27.83 \\
	Gaia DR2 6777983001871556096	&	0.342442	& 0.6860	&	4.60 & 14.006	&	27.83 \\
Gaia DR2 5810253924362115328	&	0.348502	& 0.6715	&	3.75& 13.331	&	57.38 \\
	CRTS J153119.1+423058	&	0.349549	& 0.6855	&	1.64 & 13.194	&	54.15 \\
	Gaia DR2 6600185988767577984	&	0.352454	& 0.6861	&	3.90 & 13.522	&	27.83 \\
	BPS BS 16924-0010	&	0.352996	& 0.6848	&	1.71 & 13.575	&	26.75 \\ %
	UCAC4 539-052988	&	0.360306	& 0.6800	&	2.64 & 13.510	&	25.88 \\
  Gaia DR2 5807743739318986752	&	0.368218	& 0.6863	&	2.93 & 13.894	&	27.90 \\
	CRTS J033556.8-044542	&	0.371555	& 0.6861	&	2.78 & 13.963	&	25.67 \\
	IY Eri	&	0.375022	& 0.6858	&	2.60& 10.832	&	55.21 \\
	Gaia DR2 5307694509672719872	&	0.376984	& 0.6871	&	3.06& 13.804	&	52.35 \\
\hline
\end{tabular}
\tablefoot{
Periods $P_1$ and frequency ratios $f_{68}/f_1$ are from \citet{Benko2023}.
The amplitude ratios $A_{68}/A_{59}$ were calculated by using table 3 of \citet{Benko2023}.
The TESS average brightness $T_{\mathrm{mag}}$ is taken from the TIC catalogue version 8.2 \citep{TIC_8.2}, $\Delta t$ shows the length of the data series.}
\end{table*}  

\begin{table*}
\caption{\label{Tab:param_RRc_K2}Non-Blazhko RRc stars of the K2 mission showing $f_{68}$ additional frequencies.
}
\centering
\begin{tabular}{llcclc}
\hline\hline
\noalign{\smallskip}
 EPIC ID & $P_1$ & $f_{68}/f_1$ & $A_{68}/A_{59}$ & $K_{\mathrm{p}}$ & $\Delta t$  \\
 & (d) &  & & (mag) &    (d)  \\
\hline
\noalign{\smallskip}
250034040	&	0.285430	&	0.6729	&	3.20 & 14.879	&	88.00 \\ % Moire bezavarhat!
251809817	&	0.299131	&	0.6837	&	2.87 & 19.008	&	67.07 \\
212482472	&	0.312544	&	0.6873	&	3.20 & 16.031	&	67.07  \\
212085647	&	0.318399	&	0.6851	&	3.29 & 16.933	&	79.56  \\
228710407	&	0.321891	&	0.6848	&	2.78 & 15.393	&	75.37  \\
248481034	&	0.324209	&	0.6876	&	1.32 & 17.056	&	79.68  \\
249806161	&	0.329117	&	0.6829	&	4.73 & 16.785	&	88.00  \\
246064096	&	0.334892	&	0.6835	&	2.33 & 15.198	&	78.89, 15.10\tablefootmark{$\ast$} \\
234669323	&	0.335037	&	0.6861	&	2.27 & 15.344	&	74.17  \\
\hline
\end{tabular}
\tablefoot{
The numerical values were taken from or calculated based on the work of \citet{Netzel2023}.\\
\tablefoottext{$\ast$}{Observed in two non-consecutive campaigns (C12 and C19).}
}
\end{table*} 
\end{appendix}

\end{document}